\documentclass[%
preprint,
nofootinbib, nobibnotes,
amsmath,amssymb, aps, prd,
longbibliography
]{revtex4-1}

\usepackage{graphicx}
\usepackage{bm}

\def\be{\begin{equation}}
\def\ee{\end{equation}}
\def\bea{\begin{eqnarray}}
\def\eea{\end{eqnarray}}

\def\ptl{\partial}
\usepackage{color}

\begin{document}

\centerline{\large{ \bf Evidence of time evolution in quantum
gravity}}
\bigskip
\bigskip
\centerline{S.L. Cherkas $^{a,}\footnote{Corresponding author}$,
V.L. Kalashnikov$^b$}
\bigskip
\bigskip
\centerline{\small $^a$Institute for Nuclear Problems, Bobruiskaya
11, Minsk 220030, Belarus} \centerline{\small Email:
cherkas@inp.bsu.by}
\medskip
\centerline{\small $^b$Dipartimento di Ingegneria dell’Informazione, Elettronica e Telecomunicazioni,}
\centerline{\small Sapienza Universit\'a di Roma, Via
Eudossiana 18, 00189 - Roma, RM, Italia} \centerline{\small Email:
vladimir.kalashnikov@uniroma1.it}
\bigskip
\centerline{\small $20^{th}$ of March, 2020}
\bigskip

{\small We argue that the problem of time is not a crucial issue inherent in the quantum picture of the universe evolution. On the minisuperspace model example with the massless scalar field, we demonstrate four approaches to the description of quantum evolution, which give similar results explicitly. The relevance of these approaches to building a quantum theory of gravity is discussed.}


\section{Introduction}

Usually, some crucial theoretical problems were self-created in some sense, and then these issues were solved successfully during some period. An example could be the spin crisis problem, which had been stating about 30 years \cite{RevModPhys}. The problem of time \cite{Kuchar1991,Isham1993,Shest2004,rovelli2009,anderson2010} holds a relatively long time from \cite{DeWitt67} and was related closely with the variety of the points of view in a gravity quantization. The root of this issue is the gauge invariance of the general relativity. Such invariance allows choosing the equivalent time parametrizations, and one may suspect that the time is an ``illusion.''

On the other hand, astrophysical data demonstrate undoubtedly the time evolution of the universe. The modern trends in the interpretation of quantum mechanics (e.g., see \cite{wallace}) suggest that all the phenomena, including the universe itself, are generally quantum. Thus, the time evolution in the frameworks of quantum cosmology has to been explained.

Although “eternity” and “time” are two sides of one coin \cite{Proclus}, all observations are performed in the time. Thus, the time should be put into a theory, in any case, to confront theory with the observations. However, sometimes, it could be useful to think in terms of eternity for the development of theoretical concepts \textit{sub specie aeternitatis}.

The complexity of the full system of the equations of gravity does not prevent to consider this problem on an example 
of the so-called minisuperspace models \cite{Bojowald}, which are extremely simple but have the Hamiltonian constraint like that in the general case.

Here we show that the problem of time does not prevent to calculate the time-dependent mean 
values, which could be, in principle, compared with the observations.

\section{Classical picture}

As it is well-known, there is no problem with defining the time in the classical theory because it implies that if 
an observer has some particular clock, she can choose a gauge corresponding to this clock.

Let us consider action for gravity and a real massless scalar field  $\phi$:
\bea
S=\frac{1}{16\pi G}\int R  \sqrt{-g}\,d^4x
+\frac{1}{2}\int\ptl_\mu\phi\,
g^{\mu\nu}\ptl_\nu\phi\sqrt{-g}\,d^4x,
\label{lag}
\eea
where $R$ is a scalar curvature.

We restrict the consideration by the uniform, isotropic and  flat universe
\be
ds^2=g_{\mu\nu}dx^\mu dx^\nu=a^2(N^2d\eta^2-d^2\bm r),
\ee
where a scale factor $a$ and a lapse function $N$ depend on a conformal time $\eta$ only. Under these
conditions, the action (\ref{lag}) becomes
\be
S=\frac{1}{2}\int\frac{1}{N}\left(-{M_p^2}a^{\prime
2}+a^2\phi^{\prime 2}\right)d \eta,
\label{lag1}
\ee
where the reduced Planck mass $M_p=\sqrt{\frac{3}{4\pi G}}$ is used\footnote{The scale factor $a$ in (\ref{lag1}) becomes dimensional because it corresponds, in fact, to $a V^{3/2}$, where $V$
is the volume of spatial integration in (\ref{lag}).}, which  will be set to unity in the further consideration for
simplicity.

The action (\ref{lag1}) in the generalized form looks as
\be
S=\int \left(-p_a a^\prime+\pi_\phi\phi^\prime- N\left(-\frac{1}{2
}p_a^2+\frac{\pi_\phi^2}{2a^2}\right)\right)d \eta, \label{L1}
\ee
which turns to (\ref{lag1}) after variation on $\pi_\phi$ and
$p_a$.  The explicit expression for the
Hamiltonian follows from (\ref{L1}):

\be
H=N\left(-\frac{1}{2}p_a^2+\frac{\pi_\phi^2}{2a^2}\right),
\ee
which is also the Hamiltonian constraint
\be
 \Phi_1=-\frac{1}{2
}p_a^2+\frac{\pi_\phi^2}{2a^2}=0,
\label{hamcon}
\ee
due to $\frac{\delta S}{\delta N}=0$.

Time evolution of an arbitrary quantity is expressed through the Poisson brackets
\be
\frac{dA}{d\eta}= \frac{\ptl A}{\ptl \eta}+\{H,A\},
\ee
which read as

\be
 \{A,B\}=\frac{\ptl A}{\ptl \pi_\phi}\frac{\ptl
B}{\ptl \phi}-\frac{\ptl A}{\ptl \phi}\frac{\ptl B}{\ptl
\pi_\phi}-\frac{\ptl A}{\ptl p_a}\frac{\ptl B}{\ptl a}+\frac{\ptl
A}{\ptl a}\frac{\ptl A}{\ptl p_a}.
\ee

The full system of the equations of motion has the form:
 \bea
 \pi_\phi^\prime=-\frac{\ptl H}{\ptl\phi }=0,~~\Longrightarrow~~\pi_\phi=const\equiv k,\nonumber\\
 \phi^\prime=\frac{\ptl H}{\ptl \pi_\phi}=\frac{k}{a^2},~~~
a^\prime=-\frac{\ptl H}{\ptl p_a}=p_a,~~~
 p_a^\prime=\frac{\ptl H}{\ptl a}=-\frac{k^2}{a^3}.
 \eea
The solution of the equations of motion is
\be
a=\sqrt{2|\pi_\phi| \eta},
~~~~~~~~\phi=\frac{\pi_\phi}{2|\pi_\phi|}\ln\eta +const.
\label{at}
\ee

According to Eq. (\ref{at}), a gauge fixing condition
\be
\Phi_2=a-\sqrt{2|\pi_\phi| \eta}=0,
\label{f2}
\ee
which conserves in time, can be introduced
in addition to the constraint $\Phi_1$. 

One can see that there is an explicit time evolution under some particular gauge fixing. Moreover, for this simple example, the system could be reduced to a single degree of freedom \cite{Barv2014,Kamen2019}.

Let us take $\pi_\phi$ and $\phi$ as the physical variables, then
$a$ and $p_a$ have to be excluded by the constraints (\ref{hamcon}),(\ref{f2}). Substituting
$p_a$, $a^\prime$ and $a$ into (\ref{lag1}) results in
\be
S=\int \left(\pi_\phi\phi^\prime
-H_{phys}(\phi,\pi_\phi,\eta)\right)d\eta,
\label{L2}
\ee
where
\be
H_{phys}(\phi,\pi_\phi,\eta)=p_a
a^\prime=\frac{|\pi_\phi|}{2\,\eta}.
\label{hph}
\ee

\section{Quantum pictures with time}

\subsection{Schr\"{o}dinger equation with a physical Hamiltonian}
\label{phys}

The most simple and straightforward way to the description of the quantum evolution is based on the Schr\"{o}odinger equation \cite{Barv2014,Kamen2019}
\be
i\ptl_\eta \Psi=\hat H_{phys}\Psi
\label{sh}
\ee
with a physical Hamiltonian (\ref{hph}). In the momentum
representation, the operators become
\be
\hat \pi_\phi=k, ~~~~~\hat \phi=i\frac{\ptl}{\ptl k}.
\ee

The solution of Eq. (\ref{sh}) is
\be
\Psi(k,\eta)=C(k)|2 k\eta|^{-i|k|/2}e^{i|k|/2},
\label{wave}
\ee
where $C(k)$ is a momentum wave packet.  An arbitrary operator $\hat A$
 build from $\hat
\phi=i\frac{\ptl}{\ptl k}$ and $a=\sqrt{2|k|\eta}$ is, in fact, the function of $\eta$, $k$, and $i\frac{\ptl}{\ptl k}$.
Using the wave function (\ref{wave}) allows calculating the mean value
\be
<C|\hat A|C>=\int \Psi^*(k,\eta)\hat A\,\Psi(k,\eta)dk.
\ee
\begin{figure}
  \includegraphics[width=10cm]{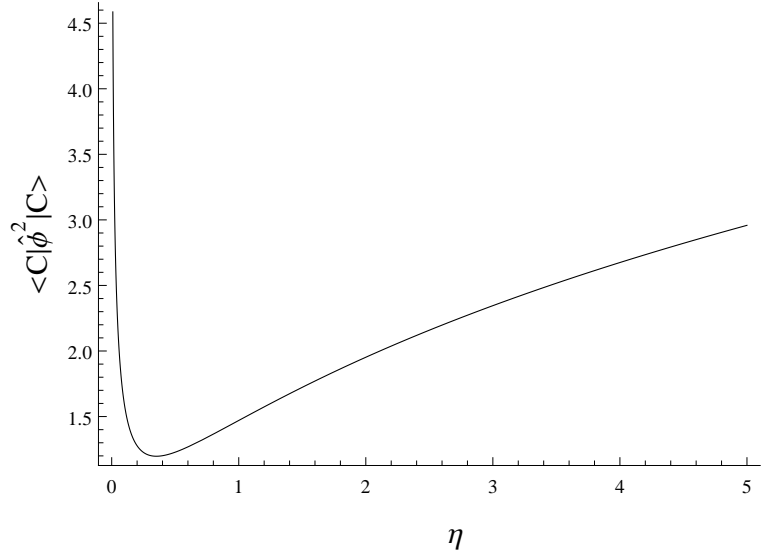}\\
  \caption{The mean value of the square of the scalar field over the wave packet (\ref{pak}).
   }\label{fig1}.
\end{figure}
Since the base wave functions $\psi_k=|2 k\eta|^{-i|k|/2}e^{i|k|/2}$ contain the module of $k$, a singularity may arise at $k=0$ if $\hat A$ contains the degrees of the differential operator $\frac{\ptl}{\ptl k}$. That could violate hermicity. To avoid this, the wave packet $C(k)$ should turn to zero at $k=0$. For instance, it could be taken in the form of the Gaussian function multiplied by $k^2$
\be
C(k)=\frac{4 \sigma^{5}}{3\sqrt \pi}k^2\exp\left(-\frac{k^2}{2\sigma^2}\right).
\label{pak}
\ee

Let us come to the calculation of the concrete mean values taking $\sigma=1$ for simplicity. Mean values of the operators $\hat \phi^2$ and $a$ are
\bea
<C|a|C>=\frac{4}{3} \sqrt{\frac{2}{\pi
}}\sqrt{\eta}\int_{-\infty}^\infty e^{-k^2} k^{9/2} dk=\frac{4}{3}
\sqrt{\frac{2}{\pi
}}\Gamma(11/4)\sqrt{\eta},~~~~~~~~~~~~~~~~~~~\label{resa}\\
<C|\hat\phi^2|C>=\frac{1}{3\sqrt{\pi}}\int_{-\infty}^\infty
e^{-k^2}\biggl(-4 k^6+k^4 \left(20+\ln ^22+\ln (\eta  \left|
k\right| ) \ln (4 \eta \left| k\right| )\right)\nonumber-\\8 k^2+2
i \left| k\right| ^3 \left(-2 k^2 \ln (2 \eta \left| k\right| )+4
\ln (\eta  \left| k\right| )+4\ln
   2+1\right)\biggr)dk=\nonumber\\
   \frac{1}{12} \ln \eta  (3 \ln \eta -3 \gamma +8)+\frac{\pi ^2}{32}+\frac{\gamma ^2}{16}-\frac{\gamma
   }{3}+\frac{4}{3},~~~~~~~~~~~~
   \label{resphi2}
\eea
 where $\Gamma$ is the Gamma function, and $\gamma$ is the Euler constant. Let us note that the imaginary part in (\ref{resphi2}) disappears after integration on $k$ due to hermicity of $\hat \phi$. Fig. \ref{fig1} demonstrates the infiniteness of the mean-square value of $\phi$ at $\eta=0$, then it decreases and begins to increase finally.

Another more complicated example is
\bea
<C|\hat \phi^2\,a+a\,\hat \phi^2|C>= \frac{1}{3072}\biggl(
   16 \ln\eta  \bigl(\ln\eta (4 \ln\eta  (3 \ln\eta -6 \gamma +16)+9 \pi ^2\nonumber\\+6 \gamma  (3 \gamma -16)+384)-9 \gamma  \pi ^2+24 \pi ^2
   -6 \gamma  (64+(\gamma -8)
   \gamma )+800\bigr)\nonumber\\+224 \zeta (3) (-6 \ln\eta +3 \gamma -8)+21 \pi ^4+12 \gamma  (3 \gamma -16) \pi ^2+768 \pi ^2\nonumber\\
   +4 \gamma  \bigl(\gamma  (384+\gamma  (3 \gamma
   -32))-1600\bigr)+16640\biggr),
   \label{corr}
   \eea
where $\zeta (x)$ is the Zeta-function.

\subsection{Time evolution from the WDW equation}
\label{wdw}
The problem of time began from the discussion of the WDW equation \cite{Wheel67,DeWitt67,Wheel87,Barv2014,Shest2018}, which is a ``workhorse'' of quantum cosmology and a ``mathematical implementation of eternity.'' It is often stated that the WDW equation does not contain time explicitly. Indeed, it is true. Then, it is usually stated that the WDW equation forbids time evolution. Certainly, it is wrong if one considers a full quantum picture, including gauge fixing and evaluation of the mean values of the operators. The point is that the WDW equation has to be supplemented by the scalar product.

 Let us to introduce the variable $\alpha=\ln a$ and perform the canonical quantization
 \be
[\hat p_\alpha,\alpha]=i,~~~~~[\hat \pi_\phi,\phi]=-i
\label{com}
 \ee
of the constraint $\Phi_1=0$. That results in the WDW equation
\be
\left(\frac{\ptl^2}{\ptl\alpha^2}-\frac{\ptl^2}{\ptl\phi^2}\right)\Psi(\alpha,\phi)=0
\label{WDW1}
\ee
 of the Klein-Gordon type.

 Scalar products for the Klein-Gordon equation are discussed in \cite{mostf}, where the ``current'' and ``density'' products were proposed. Here we will use only the scalar product of the ``current'' type:
\be
<\Psi|\Psi>=i\int \left(\Psi^*(\alpha,\phi)\frac{\ptl }{\ptl
\alpha}\Psi(\alpha,\phi)-\Psi(\alpha,\phi)\frac{\ptl}{\ptl
\alpha}\Psi^*(\alpha,\phi)\right)\biggr|_{\,\alpha=\alpha_0}d\phi,
\label{kl}
\ee
including the hyperplane $\alpha=\alpha_0$.
For a mean value of some operator, the following formula has been
introduced \cite{mostf}
\be
  <\Psi|\hat A|\Psi>=i \int \biggl( \Psi^* {\hat D^{1/4}}\hat A\,{\hat
  D^{-1/4}}\frac{\ptl \Psi}{\ptl
\alpha }-\left(\frac{\ptl
\Psi^*}{\ptl \alpha }\right){\hat D^{-1/4}}\hat A\,{\hat
D^{1/4}}\,\Psi\biggr)\biggr|_{\,\alpha=\alpha_0} d\phi
,\label{mean3}
\ee
where $\hat D=-\frac{\ptl^2}{\ptl\phi^2}$. In the
momentum representation $\hat \pi_\phi=k$, $\hat \phi=i\frac{\ptl
}{\ptl k}$, the WDW equation (\ref{WDW1}) reads as
\be
\left(\frac{\ptl^2}{\ptl\alpha^2}+k^2\right)\psi(\alpha,k)=0,
\ee
and, as a result of $\hat D^{1/2}=|k|$, the scalar product (\ref{mean3}) takes the form:
\be
<\Psi|\hat A|\Psi>=i \int C^*(k)e^{i|k|\alpha}\hat A
e^{-i|k|\alpha}C(k)\biggr|_{\,\alpha=\alpha_0}dk,
\label{meanf}
\ee
where
\be \Psi(\alpha,\phi)=\int e^{ik\phi}
\psi(\alpha,k)dk=\int
\frac{e^{ik\phi-i|k|\alpha}}{\sqrt{2|k|}}C(k)dk
\ee is taken. To introduce the time evolution into this picture, one
has to choose a time-dependent integration plane in (\ref{meanf})
instead of $\alpha=\alpha_0$ by writing
$\alpha=\frac{1}{2}\ln\left(2|k|\eta\right)$ according to
(\ref{f2}), i.e., to $\Phi_2=0$.

However, if the
operator $\hat A(\alpha,k,i\frac{\ptl}{\ptl
\alpha},i\frac{\ptl}{\ptl k})$ contains differentiations
$\frac{\ptl}{\ptl k}$ or $\frac{\ptl}{\ptl \alpha}$, hermicity
could be lost. To prevent this, let us rewrite (\ref{mean3}),
(\ref{meanf}) in the form of
\bea
<\psi|\hat A|\psi>= \int \psi^*(\alpha,k)\biggl(|k|^{1/2}\hat
A|k|^{-1/2}\delta(\alpha -\frac{1}{2}\ln(2|k|\eta))\hat
p_\alpha+~~~~~~~~\nonumber\\\hat p_\alpha\delta(\alpha
-\frac{1}{2}\ln(2|k|\eta))|k|^{-1/2}\hat
A|k|^{1/2}\biggr)\psi(\alpha,k) d \alpha dk,
\label{sc2}
\eea
where $p_\alpha=i\frac{\ptl}{\ptl \alpha}$ and hermicity of $\hat A$ relatively $\alpha,k$ variables are implied. In this case, no problem with hermicity arises if one takes the functions $\psi(\alpha,k)$ tending to zero at $\alpha\rightarrow \pm \infty$ to provide the throwing over the differential operators $\ptl/\ptl \alpha$ by the integration by parts. The functions
$\psi(\alpha,k)=\frac{e^{-i|k|\alpha}}{\sqrt{2|k|}}C(k)$ do not
possess such a property. Thus, we shall take the functions
\be
\psi(\alpha,k)=\frac{e^{-i|k|\alpha-\alpha^2/\Delta}}{\sqrt{2|k|}}C(k)
\label{ftest}
\ee
in the intermediate calculations and, then, after integration over
$\alpha$, tend $\Delta$ to infinity. Performing the concrete calculations with the above
wave packet (\ref{pak}), we
again obtain the same values for (\ref{resa}) and (\ref{resphi2}).
As for the mean value (\ref{corr}) of subsection \ref{phys}, we cannot compare it using this picture because the particular operator ordering $a \hat \phi^2+\hat \phi^2 a$ has been used in (\ref{corr}), but here the operators $a=\exp{\alpha}$ and $\hat \phi$ commute formally implying an existence of some intrinsic automatic ordering.

\subsection{An evolution from the WDW using the Grassman variables}
\label{wdwgr}

Another version with the anticommutative variables could be proposed in the form
\bea
<\psi|A|\psi>= \int
\psi^*(\alpha,k)\exp\biggl(i\lambda\bigl(\alpha
-\frac{1}{2}\ln(2|k|\eta)\bigr)+\bar \theta\theta
\hat p_\alpha+~~~~~~~~~~~\nonumber\\\frac{1}{2}\bar \chi\chi\left(|k|^{-i/2} \hat A\,|k|^{i/2}+|k|^{i/2} \hat A\,|k|^{-i/2}\right)\biggr)\psi(\alpha,k)d\lambda
d\alpha  dk d \theta d\bar\theta d \chi d\bar \chi,
\label{sc22}
\eea
where the anticommutating Grassman variables $\theta_i=(\theta,\chi)$, $\bar \theta_i=(\bar\theta,\bar\chi)$ have the following
properties: $\theta_i\theta_j+\theta_j\theta_i=0$, $\int d\theta_i=0$, $\int \theta_i d\theta_i=1$, $(\bar
\theta_i)^*=\theta_i$, $(\bar \theta_i\theta_j)^*=\bar \theta_j\theta_i$. Again, for reasons of hermicity, we take the functions 
(\ref{ftest}) and then tend $\Delta$ to
infinity. For the practical calculations, it is convenient to separate
the expression in the exponent of Eq. (\ref{sc22}) into two parts
$M=i\lambda\left(\alpha
-\frac{1}{2}\ln\left(2|k|\eta\right)\right)$, and $N=\bar \theta\theta
\hat p_\alpha+\frac{1}{2}\bar \chi\chi\left(|k|^{-i/2} \hat A|k|^{i/2}+|k|^{i/2} \hat A|k|^{-i/2}\right)$ for using  the
formula \cite{Kimura2017}
\be
\exp\left(\hat M+\hat N\right)=\left(1+\sum_{m=1}^{\infty}\frac{\hat X_m}
{m!}\right)\exp{\hat M},
\label{sum}
\ee
where $\hat X_m$ is set recurscively as  $\hat X_1=N$ and $\hat X_m=\hat N \hat X_{m-1}+[\hat M,\hat X_{m-1}] $. In fact, it is sufficient to take only finite number of terms in 
a sum of Eq. (\ref{sum}).

\subsection{Quasi-Heisenberg picture}
\label{quas}

Another approach to consider the time evolution is to take classical equations of motion and then quantize them, i.e., write ``hat'' over every quantity \cite{Cher2005,Cher2012,Cher2013,Cher2017}. The operator equations of motion take the form:
\be
\hat \phi^{\prime\prime}+2\hat \alpha^\prime\hat
\phi^\prime=0,~~~~~~~~~~~ \hat \alpha^{\prime\prime}+\hat
\alpha^{\prime 2}+\hat \phi^{\prime 2}=0.
\label{eqop}
\ee
One needs to find the commutation relations of the operators $\hat \alpha(\eta)$, $\hat \phi(\eta)$. The problem was solved by Dirac, who has introduced the ``Dirac brackets'' for a system with constraints postulating that commutator relations of the operators have to be analogous to the Dirac brackets. However, it is not always possible to find an operator realization of this commutator relations. The quasi-Heisenberg picture suggests to find an operator realization only at the initial moment and then allow operators to evolve according to the equations of motion. The initial conditions for operators could be taken in the form
\be
\hat \alpha(0)=\alpha_0, ~~~ \hat
\alpha^\prime(0)=e^{-2\alpha_0}|k|,~~~\hat
\phi(0)=i\frac{\ptl}{\ptl k}, ~~~\hat
\phi^\prime(0)=e^{-2\alpha_0}k.
\label{in}
\ee

The solution of the operator equations of motion (\ref{eqop}) with the
initial conditions (\ref{in}) is
\be
\alpha(\eta)=\alpha_0+\frac{1}{2}\ln\left(1+2|k|\eta\,
e^{-2\alpha_0}\right),~~~\hat \phi(\eta)=i\frac{\ptl}{\ptl
k}+\frac{k}{2|k|}\ln\left(1+2|k|\eta \,e^{-2\alpha_0}\right).
\ee

To built the Hilbert space, in which these quasi-Heisenberg operators act, one may use the WDW equation (\ref{WDW1}) and the scalar product (\ref{meanf}) but, at the end of mean values evaluating, the value of $\alpha_0$ should be set to minus infinity, i.e., $\alpha_0\rightarrow -\infty$, which corresponds to $a\rightarrow 0$ at $\eta=0$. Explicit calculation gives the same mean values as (\ref{resa}), (\ref{resphi2}) and (\ref{corr}).

\subsection{Evolution using the unconstraint Schr\"{o}dinger equation in the extended space}
\label{shev}

It is believed \cite{Kaku2012,Savchenko2004,Vereshkov2013} that the Grassman variables allow writing the Lagrangian without constraints at all. Here, one has two possibilities: to set a gauge imposing an additional condition as a function of  $p_a$, $a$, $\pi_\phi$, $\phi$ such as (\ref{f2}) (canonical gauge). Another alternative is to impose that condition as a function of $N$ (non-canonical gauge).  

\subsubsection{Canonical gauge}

The discussion  can be started in terms of continual integral which gives a transition amplitude from $in$  to $out$  states \cite{Feynman,Faddeev,Kaku2012}:
\be
<out|in>=
Z=\int  e^{i\int\left(\pi_{\phi}\phi^\prime
-H_{phys}(\phi,\pi_\phi)\right)d\eta} D \pi_\phi \mathcal D\phi,
\label{z00}
\ee
where $H_{phys}$ is given by (\ref{hph}).
This functional can be rewritten in the form
\be
Z=\int e^{i\int\left(\pi_{\phi}\phi^\prime -p_a
a^\prime-N\left(-\frac{1}{2}p_a^2+\frac{\pi_{\phi}^2}{2a^2}\right)\right)d\eta}\Pi_\eta
p_a\Pi_\eta \delta\bigl(a-\sqrt{2\eta|\pi_\phi|}\bigr)\mathcal D p_a
\mathcal D a\mathcal D \pi_\phi \mathcal D \phi\mathcal D N,
\label{z0}
\ee
where \cite{Faddeev} $p_a=\{\Phi_1,\Phi_2\}$ is the Faddeev-Popov determinant.  Equivalence of (\ref{z00}) and (\ref{z0})
can be checked by transition to a new integration variable $\tilde
a=a-\sqrt{2\eta|\pi_\phi|}$, and integrating on $\tilde a$, $N$,
$p_a$ in (\ref{z0}) gradually.

Using the Grassman anticommutative variables in Eq. (\ref{z0}) leads
to the form containing the unconstraint Lagrangian in the exponent
\be
Z=\int e^{i\int\bigl(\pi_{\phi}\phi^\prime -p_a
a^\prime-N\bigl(-\frac{1}{2}p_a^2+\frac{\pi_{\phi}^2}{2a^2}\bigr)-\lambda\left(a-\sqrt{2\eta|\pi_\phi|}\right)-\bar\theta\theta
p_a\bigr)d\eta}\mathcal D p_a \mathcal D a\mathcal D \pi_\phi
\mathcal D \phi\mathcal D N\mathcal D \lambda \mathcal D
\theta\mathcal D\bar \theta.
\label{final}
\ee

Eq. (\ref{final}) allows writing the Hamiltonian
\be
H=N\biggl(-\frac{1}{2}p_a^2+\frac{\pi_{\phi}^2}{2
a^2}\biggr)+\lambda\left( a-\sqrt{2\eta|
\pi_\phi|}\right)+\bar\theta\theta p_a,
\label{ham}
\ee
which, after canonical quantization, could be used to describe evolution as in both Schr\"{o}dinger and Heisenberg pictures.

\subsubsection{Non-canonical gauge}

Let us remind, how the Faddeev-Popov determinant appears in non-canonical gauge. The action (\ref{lag1}) is invariant relatively the infinitesimal gauge transformations:
\bea
\tilde a=a+\delta a=a+\varepsilon\, a^\prime,\\
\tilde \phi=\phi+\delta \phi=\phi+\varepsilon\, \phi^\prime,\\
\tilde N=N+\delta N=N+(N\varepsilon)^\prime,\label{NN}
\eea
where $\varepsilon$ is an infinitesimal  function of time. When one sets a non-canonical gauge condition in the form $F(N)=0$, the functional integral, including a gauge fixing multiplier with the Dirac $\delta$-function, becomes \cite{Kaku2012}
\be
Z=\int e^{i\int\left(\pi_{\phi}\phi^\prime -p_a
a^\prime-N\left(-\frac{1}{2}p_a^2+\frac{\pi_{\phi}^2}{2a^2}\right)\right)d\eta}\Pi_\eta
\frac{\delta F}{\delta \varepsilon}\Pi_\eta \delta(F)\mathcal D
p_a \mathcal D a\mathcal D \pi_\phi \mathcal D \phi\mathcal D N,
\ee
where again the Faddev-Popov determinant $\Delta_{FP}=\frac{\delta
F}{\delta \varepsilon}$ has been introduced \cite{Kaku2012}. In the
particular case $F=N-1$, it follows from (\ref{NN}) that the
determinant is $\Delta_{FP}=\frac{\delta N}{\delta \varepsilon}=N^\prime+N\frac{\ptl}{\ptl \eta}$.
Using the Grassman variables 
raises the determinant into
an exponent
\bea
Z=i\int e^{i\int\left(\pi_{\phi}\phi^\prime -p_a
a^\prime-N\left(-\frac{1}{2}p_a^2+\frac{\pi_{\phi}^2}{2a^2}\right)-\lambda(N-1)-
N^\prime\bar\theta \theta-N\bar
\theta\theta^\prime\right)d\eta}\mathcal D p_a \mathcal D
a\mathcal D \pi_\phi \mathcal D\phi\mathcal D \lambda\mathcal D
N\mathcal D \theta\mathcal D\bar \theta\nonumber\\= \int
e^{i\int\left(\pi_{\phi}\phi^\prime -p_a
a^\prime-\left(-\frac{1}{2}p_a^2+\frac{\pi_{\phi}^2}{2a^2}\right)-
\bar \theta\theta^\prime\right)d\eta}\mathcal D p_a \mathcal D
a\mathcal D \pi_\phi D \phi\mathcal D \theta\mathcal D\bar \theta.
~~~~~~\label{actg}
\eea
An expression in the exponent of Eq. (\ref{actg}) could be treated as Lagrangian, but it cannot be put into the generalized Hamiltonian form, because velocity $\theta^\prime$  cannot be expressed through a momentum. In this relation, an interesting trick has been suggested \cite{Savchenko2004}: to take the gauge condition $N^\prime=0$, instead of $N=1$. For this new gauge, it follows from (\ref{NN}) that
\be
\delta F=\delta N^\prime=(N \varepsilon)^{\prime\prime},
\ee
and
\bea
Z=\int e^{i\int\left(\pi_{\phi}\phi^\prime -p_a
a^\prime-N\left(-\frac{1}{2}p_a^2+\frac{\pi_{\phi}^2}{2a^2}\right)-\lambda
N^\prime- \bar\theta (N \theta)^{\prime\prime}\right)d\eta}
\mathcal D p_a \mathcal D a\mathcal D \pi_\phi \mathcal
D\phi\mathcal D N\mathcal D \lambda\mathcal D \theta\mathcal D\bar
\theta. ~~~~\label{actg1}
\eea
 The unconstraint Lagrangian is written from Eq. (\ref{actg1})  as
\be
L=\pi_{\phi}\phi^\prime -p_a
a^\prime-N\left(-\frac{1}{2}p_a^2+\frac{\pi_{\phi}^2}{2a^2}\right)-\lambda
N^\prime+ \bar\theta^\prime (N \theta)^\prime,
\label{lagun}
\ee

For the momentums of the Grassman variables and $N$, one has from
(\ref{lagun})
\be
\pi_\theta=-\frac{\ptl L}{\ptl \theta^\prime}=N\bar
\theta^\prime,~~~\pi_{\bar \theta}=\frac{\ptl L}{\ptl \bar
\theta^\prime}=N^\prime\theta+N\theta^\prime,~~~~p_N=\frac{\ptl
L}{\ptl N^\prime}=-\lambda+\bar \theta^\prime\theta,
\ee
where, as usual, the left derivative over the Grassman variables $\frac{\ptl}{\ptl
\theta}\left(\theta f(\bar\theta)\right)=f(\bar \theta)$ is implied. 
The Lagrangian (\ref{lagun}), rewritten in terms of momentums acquires the form 
\be
L=\pi_{\phi}\phi^\prime -p_a
a^\prime+p_N\,N^\prime+\bar \theta^\prime\pi_{\bar \theta}+\pi_\theta\theta^\prime-N\left(-\frac{1}{2}p_a^2+\frac{\pi_{\phi}^2}{2a^2}\right)-\frac{1}{N}\pi_
\theta\pi_{\bar \theta}.
\ee
This means that the corresponding Hamiltonian is 
\be
H=N\left(-\frac{1}{2}p_a^2+\frac{\pi_{\phi}^2}{2a^2}\right)+\frac{1}{N}\pi_\theta\pi_{\bar \theta}.
\label{ham2}
\ee
Thus, two Hamiltonians (\ref{ham}), (\ref{ham2}), which drive unconstraint dynamics, have been obtained. The first one is time-dependent and contains the Grassman variables as parameters. The second one is time-independent and implies the time dynamics of the Grassman variables \cite{Savchenko2004}.
Further, we will consider only the Hamiltonian (\ref{ham2}), because this timeless Hamiltonian seems more perspective in the general gravity quantization. 
Opposite to commutation relation (\ref{com}), the anticommutation relation have to be introduced for the Grassman variables

\be
\{\pi_\theta,\theta\}=-i,~~~~\{\pi_{\bar \theta},\bar \theta\}=-i.
\ee
In
the particular representation $\alpha=\ln a$, $\hat p_\alpha=i\frac{\ptl}{\ptl
\alpha}$, $\hat \phi=i\frac{\ptl}{\ptl k}$, $\hat \pi_\phi=k$, $\hat \pi_\theta=-i\frac{\ptl}{\ptl
\theta}$, $\hat \pi_{\bar \theta}=-i\frac{\ptl}{\ptl
\bar \theta}$, the Schr\"{o}dinger
equation reads as
\be
i\frac{\ptl}{\ptl \eta}\psi=\left(\frac{N}{2}e^{-2\alpha}\left(\frac{\ptl^2}{\ptl \alpha^2}+k^2
\right)-\frac{1}{N}\frac{\ptl}{\ptl \theta}\frac{\ptl}{\ptl \bar \theta}\right)\psi,
\label{shr}
\ee
where the operator ordering in the form of the two-dimensional Laplacian has been used.

 It should be supplemented by the scalar
product
\be
<\psi|\psi>=\int \psi^*(\eta,N,k,\alpha,\bar \theta
,\theta) \psi(\eta,N,k,\alpha,\bar \theta ,\theta)e^{2\alpha} d \alpha dk d
N  d  \theta d \bar \theta,
\ee
where the measure $e^{2\alpha}$ arises due to hermicity requirement \cite{Dew,Faddeev}.
This measure is a consequence of a minisuperspace metric if the Hamiltonian is written in the form 
$H=\frac{N}{2} g^{ij}p_i\,p_j+\frac{1}{N}\pi_\theta\pi_{\bar \theta}$ with $ p_i\equiv\{\alpha,\phi\}$, $g^{ij}=\mbox{diag}\{-e^{-2\alpha},e^{-2\alpha}\}$. Thus, 
the measure takes the form
$\sqrt{|\det{g_{ij}}|}=e^{2\alpha}$ \cite{Dew}.

 One of the particular formal solutions of the equation (\ref{shr}) could be
written as
\be
\psi(\eta,N,k,\alpha,\bar\theta,\theta)=(\bar\theta+\theta)\psi_1(\eta,N,k,\alpha)+i(\bar \theta-\theta)\psi_2(\eta,N,k,\alpha),
\ee
where the functions  $\psi_1$ and $\psi_2$ satisfy the equation 
\be
i\frac{\ptl}{\ptl \eta}\psi_{1,2}=\hat H_0\psi_{1,2},
\label{shr0}
\ee
where $\hat H_0=\frac{N}{2}e^{-2\alpha}\left(\frac{\ptl^2}{\ptl \alpha^2}+k^2
\right)$.
Then, the scalar product reduces to
\be
<\psi|\psi>=2i\int \left(\psi_2^*\psi_1-\psi_1^*\psi_2\right)e^{2\alpha} d \alpha dk d
N. 
\label{norm}
\ee
\begin{table}
\caption{Comparison of the mean values calculated by the different methods. Capital letters denote the section of a method. A plus implies that the values obtained by the different methods coincide. Crosses of two types in a circle mean that the values obtained at least by two different methods coincide.
}
\label{tab}
\begin{tabular}{llllll}
\hline
Method & A & B & C & D & E\\
\hline
$a$ & + & + & + & + &  \\
$a^2$ &+&+ &+ & +&+\\
$\hat \phi^2$ &+ &+ &+&+ & \\
$\hat \phi^4$ &$\oplus$ &$\otimes$ &$\otimes$& $\oplus$&\\
$\hat \phi^6$ &$\oplus$ & & &$\oplus$ & \\
$a\hat \phi^2 +\hat\phi^2 a~~$ &$\oplus$ & & &$\oplus$& \\\hline
\end{tabular}
\vspace*{-2pt}
\end{table}
\begin{figure}[ht]
  \includegraphics[width=10cm]{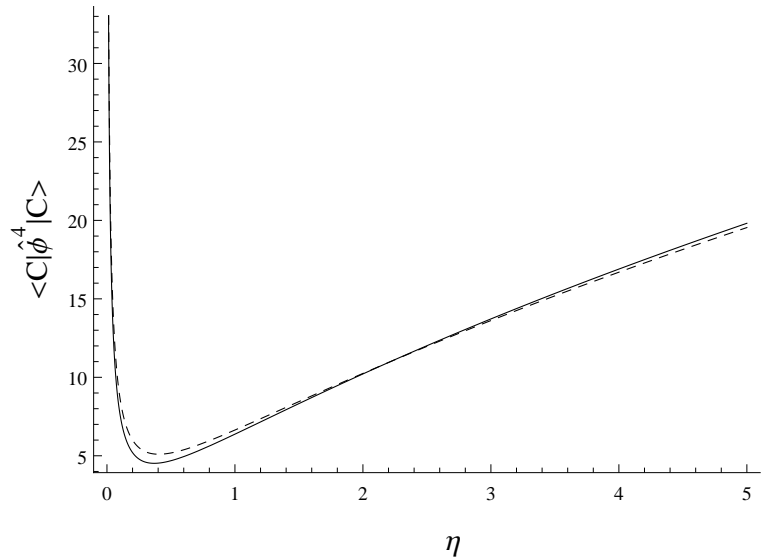}\\
  \caption{The mean value of $\hat \phi^4$ over the wave packet (\ref{pak}) for the methods A,D- solid line and methods B,C- dashed line. 
   }\label{fig2}.
\end{figure}
\begin{figure}[th]
  \includegraphics[width=6cm]{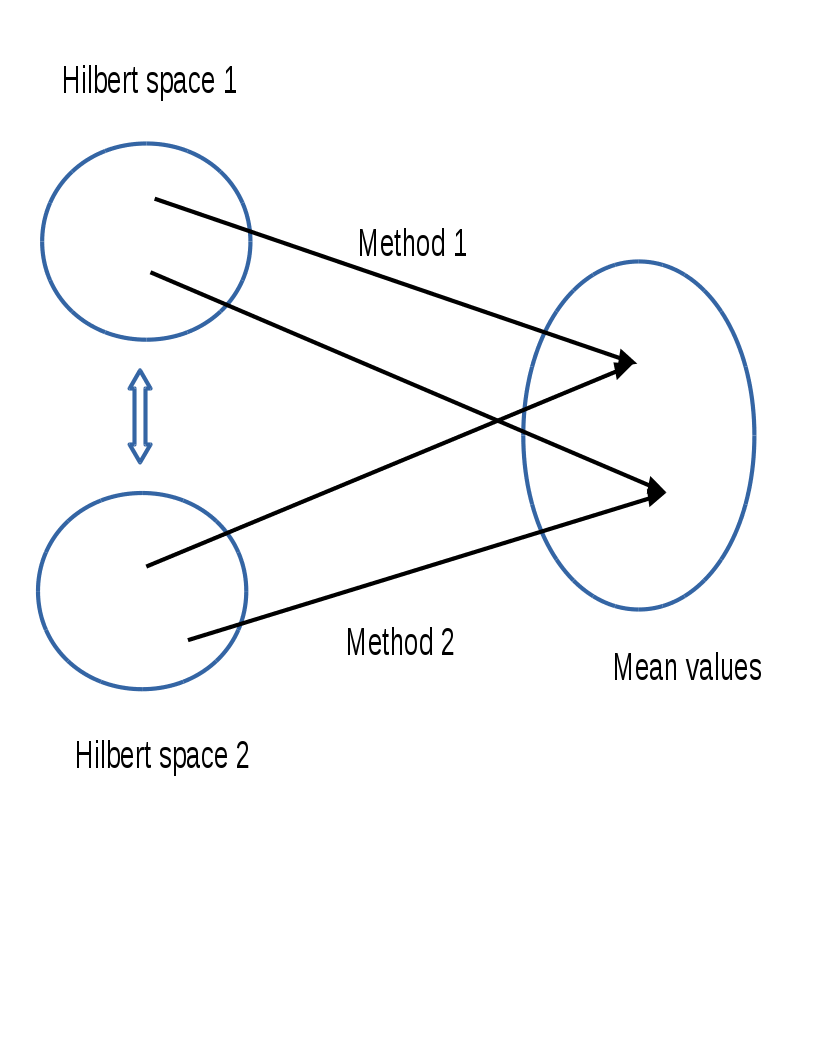}
  \vspace{-2cm}
  \caption{The illustration that different methods could have different Hilbert spaces for producing the same set of the mean values for the arbitrarily given operators. Still, there should be correspondence between the state $C(k)$ of the Hilbert space $1$ and the state $\tilde C(k)$ of the Hilbert space $2$ producing the same mean values. 
   }
   \label{fig3}.
  \end{figure}
To obtain the mean values close to that given by the previous methods, where Klein-Gordon scalar product is used, let us take the functions $\psi_1$, $\psi_2$ in the form
\be
\psi_2=e^{-i\hat H_0\eta}\psi_0(\alpha,k),~~~~\psi_1=e^{-i\hat H_0\eta}\frac{\ptl}{\ptl \alpha}
\psi_0(\alpha,k),
\label{sols}
\ee
where $\psi_0(\alpha,k)$  is given by (\ref{ftest}).
As one can see, at the limit $\Delta\rightarrow \infty$, the state (\ref{sols}) comes to the space of the WDW solution (see, e.g., \cite{Mont}), and the time 
evolution disappears. However, if this limit is taken after the calculating of the mean values, then the explicit time evolution can be caught.
Let us consider the mean value of  $a^2=e^{2\alpha}$ for the wave packet (\ref{pak}). For the variable $N$, we will consider a very narrow packet around the value $N=1$, i.e., simply set $N=1$ and abandon integration over $N$. 
Remaining integrations give for the normalizing multiplier  
\be
<\psi|\psi>=2i\int \left(\psi_2^*\psi_1-\psi_1^*\psi_2\right)e^{2\alpha} d \alpha dk =\frac{3 \pi  e^{\Delta /2} \sqrt{\Delta }}{2 \sqrt{2}}.
\ee

\noindent Then the mean value of $a^2$ becomes
\be
<a^2>=\frac{<\psi|e^{2\alpha}|\psi>}{<\psi|\psi>}=e^{3 \Delta /2}+\frac{8 (2 \Delta +1) \eta }{3 \sqrt{\pi } \Delta }+\frac{3 e^{-\Delta /2} \eta ^2}{\Delta }.
\label{sa2}
\ee
As one can see, three terms appear in Eq. (\ref{sa2}).
The first term is divergent at $\Delta\rightarrow \infty$, i.e., when one proceeds to the space of the WDW solutions, the evolution disappears, in a sense that this constant term dominates in (\ref{sa2}). However, if one omits this constant  term (not dependent on time) and then tends to the limit $\Delta\rightarrow \infty$, then the value $<a^2>=\frac{16 \eta }{3 \sqrt{\pi}}$ is the same as in the previous methods A,B,C,D. In the general case, for instance, under evaluation $a^4$, the other diverging terms depending on the time appear. That prevents extracting the time evolution when one proceeds from the extended space to the space of the WDW solutions. However, one could believe that some good regularization method could exist.  

\section{Discussion and possible application of the above approaches to 
the general case of  gravity quantization}

The results of the calculation of the mean values are presented in Table \ref{tab}. The mean value of $<C|a^2|C>$ turns out to be the same for all the methods considered. For the method \ref{shev}, we are not able to calculate the mean values of the other operators for two reasons: because we use the most primitive way of calculation by expanding the exponent $e^{-i\hat H_0\eta}$ in Eq. (\ref{sols}) over the degrees of $\eta$, and use the primitive regularization under transition from extended space \cite{Savchenko2004,Vereshkov2013} to the space of the WDW equation solutions. 

The methods A,B,C,D produce the same value of the operators $a$, $\phi^2$ as it is shown in Table \ref{tab}. For the mean value of $\hat \phi^4$, some difference emerges, which is shown in Fig. \ref{fig2}. It is not the uncertainty of numerical calculations because they are fully analytical and have been performed using Mathematica. However, let us emphasize that it does not mean that the different methods are nonequivalent. In the general case, as it is illustrated in Fig. \ref{fig3}, the different methods should not have the same Hilbert space when producing the same values of the different operators. Only the correspondence between these spaces should exist, i.e., these spaces have to be connected by some transformation.

In light of quantum gravity, one could guess that the method of subsection \ref{phys} is not likely to be applicable to the building of the general theory of quantum gravity. It is not possible, simply, to resolve the Hamiltonian and the momentum constraints to exclude the extra degrees of freedom in the general case.

Most of the considered methods use the time-dependent gauge condition. It seems the restrictive case for the general gravity if to demand conservation of the gauge condition in time. In fact, it is equivalent to the preliminary solution of the equations of motion for gravity. An exception is the quasi-Heisenberg picture \ref{quas}, which demands to set a gauge condition only at the initial moment of time. Thus, it seems the most perspective picture. The unconstrained Schr\"{o}dinger equation of subsection \ref{shev} also seems attractive \cite{Shest2018}. Still, it needs the invention of some regularization scheme when one proceeds from the extended space to the space of the WDW equation solutions.
One could hope that quantum computing will be applied \cite{Kocher2018,lloyd2005,ganguly2019} for a description of the quantum universe evolution in the future.

\section{Conclusion}

As one can see, the description of quantum evolution is very straightforward and unambiguous. Still, it teems with different details such as choosing a scalar product and operator ordering, which are typical for quantization of the systems with constraints \cite{ruffini2005}.
It is shown that if one wants to discuss the quantum evolution of the universe, there are no serious obstacles to this. Namely, the ``problem of time'' does not exist as a real problem preventing calculation of the time-dependent mean values.

Let us summarize the methods producing an explicit time evolution: 
i)	an time-dependent physical Hamiltonian with the excluded extra degrees of freedom,
ii)	the WDW equation with the time-dependent integration plane in the scalar product,
iii)	the quasi-Heisenberg picture quantizing the equations of motion, 
iv)	and an unconstraint Hamiltonian with the Grassman variables. 
Since the WDW equation tells nothing about the time evolution without determining the scalar product, this equation alone is only halfway to a full picture.

\par \noindent{\bf \small ACKNOWLEDGEMENTS}

S.L.C. is grateful to Dr. Tatiana Shestakova for discussions. 

\bibliography{time}

\end{document}